\documentclass[aps,prb,twocolumn,showpacs,preprintnumbers,amsmath,amssymb,superscriptaddress]{revtex4}%

\usepackage{graphicx}%
\usepackage{dcolumn}
\usepackage{amsmath}
\usepackage{color}
\usepackage{multirow}

\makeatletter
\def\btt#1{\texttt{\@backslashchar#1}}%
\DeclareRobustCommand\bblash{\btt{\@backslashchar}}%
\makeatother

\begin{document}

%\preprint{PREPRINT (\today)}

\title{Intrinsic and  structural isotope effects in Fe-based superconductors}
\author{R.~Khasanov}
 \email[Corresponding author: ]{rustem.khasanov@psi.ch}
 \affiliation{Laboratory for Muon Spin Spectroscopy, Paul Scherrer
Institute, CH-5232 Villigen PSI, Switzerland}
\author{M.~Bendele}
 \affiliation{Laboratory for Muon Spin Spectroscopy, Paul Scherrer
Institute, CH-5232 Villigen PSI, Switzerland}
 \affiliation{Physik-Institut der Universit\"{a}t Z\"{u}rich,
Winterthurerstrasse 190, CH-8057 Z\"urich, Switzerland}
\author{A.~Bussmann-Holder}
\affiliation{Max-Planck-Institut f\"ur Festk\"orperforschung, Heisenbergstrasse
1, D-70569 Stuttgart, Germany}
\author{H.~Keller}
 \affiliation{Physik-Institut der Universit\"{a}t Z\"{u}rich,
Winterthurerstrasse 190, CH-8057 Z\"urich, Switzerland}
%
%\author{R.~Khasanov}
% \email[Corresponding author: ]{rustem.khasanov@psi.ch}
% \affiliation{Laboratory for Muon Spin Spectroscopy, Paul Scherrer
%Institute, CH-5232 Villigen PSI, Switzerland}
%
%
%\author{H.~Keller}
%\affiliation{Physik-Institut der Universit\"{a}t Z\"{u}rich,
%Winterthurerstrasse 190, CH-8057 Z\"urich, Switzerland}

\begin{abstract}
The currently available results of the isotope effect on the superconducting
transition temperature $T_c$ in Fe-based high-temperature superconductors (HTS)
are highly controversial. The values of the Fe isotope effect (Fe-IE) exponent
$\alpha_{\rm Fe}$ for various families of Fe-based HTS were found to be as well
positive, as negative, or even be exceedingly larger than the BCS value
$\alpha_{\rm BCS}\equiv0.5$. Here we demonstrate that the Fe isotope
substitution causes small structural modifications which, in turn, affect
$T_c$. Upon correcting the isotope effect exponent for these structural
effects, an almost unique value of $\alpha \sim0.35-0.4$ is observed for at
least three different families of Fe-based HTS.
\end{abstract}

\maketitle

%Introduction

The isotope effect on the superconducting transition temperature $T_c$
traditionally plays an important role in identifying the superconducting
pairing mechanism. As a rule, an impact of the isotope substitution and,
consequently, an involvement of the lattice degrees of freedom into the pairing
mechanism are determined by comparing the isotope effect exponent
$\alpha=-(\Delta T_c/T_c)/(\Delta M/M)$ ($M$ is the atomic mass) with the
universal value $\alpha_{\rm BCS}\equiv0.5$ as predicted within the framework
of BCS theory of electron-phonon mediated superconductivity.

In conventional phonon mediated superconductors like simple metals, alloys,
{\it etc.}  $\alpha$, typically, ranges from 0.2 to 0.5, see, {\it e.g.},
Ref.~\onlinecite{Poole00} and references therein. The only exceptions are Ru
and Zr exhibiting zero isotope effect and PdH(D) with $\alpha_{\rm
H(D)}=-0.25$.\cite{IE-PdH} The negative isotope effect of PdH(D) is explained,
however, by the presence of strong lattice anharmonicty caused by the
double-well potential in the proton (deuteron) bond
distribution.\cite{Yussouff95} This was confirmed by neutron scattering data
where the large zero point motion of H in comparison with that of Deuterium
results in 20\% change of the lattice force constants.\cite{Rahman76}
A similar finding exists in organic superconductors where the H(D) isotope
effect changes sign as compared, {\it e.g.}, to $^{34}$S, $^{13}$C, and
$^{15}$N isotope replacements, see, {\it e.g.}, Ref.~\onlinecite{Schlueter01}
and references therein. Again, an unusually strong anharmonic lattice dynamics
are attributed to this observation.\cite{Schlueter01,Whangbo97}
The cuprate high-temperature superconductors (HTS) are characterized by a
vanishingly small but positive isotope effect exponent in optimally doped
compounds which increases in a monotonic way upon decreasing
doping.\cite{Batlogg87,Franck91Franck94,Zech94,Zhao01,
Khasanov04Khasanov04aKhasanov06Khasanov07Khasanov08,
Tallon05,IE_Bi2201-2212-2223,Khasanov08_IE-phase-diagram} For the optimally
doped cuprate HTS the smallest value of the  oxygen-isotope exponent
$\alpha_{\rm O}\simeq 0.02$ was obtained for YBa$_2$Cu$_3$O$_{7-\delta}$ and
Bi$_2$Sr$_2$Ca$_2$Cu$_3$O$_{10+\delta}$, while it reaches $\alpha_{\rm O}\simeq
0.25$ for
Bi$_2$Sr$_{1.6}$La$_{0.4}$CuO$_{6+\delta}$.\cite{Batlogg87,Franck91Franck94,
Zech94,IE_Bi2201-2212-2223,Khasanov08_IE-phase-diagram} In addition, it was
demonstrated that in underdoped materials $\alpha_{\rm O}$ exceeds
substantially the BCS limit $\alpha_{\rm
BCS}\equiv0.5$.\cite{Franck91Franck94,Zhao01,Khasanov08_IE-phase-diagram} It is
important to note here that the values of both, the oxygen and the copper
isotope exponents in cuprate HTS are {\it always} positive. Similar tendencies,
with the only few above mentioned exceptions, are realized in a case of
conventional phonon mediated superconductors.

Since the discovery of superconductivity in Fe-based compounds few attempts to
measure the isotope effect on $T_c$ in these materials were made. Currently we
are aware of four papers reporting, however, rather contradictory results.
\cite{Liu09,Shirage09,Shirage10,Khasanov10_FeSe-isotope} Liu {\it et
al.}\cite{Liu09} and Khasanov {\it et al.}\cite{Khasanov10_FeSe-isotope} have
found a {\it positive} Fe isotope effect (Fe-IE) exponent $\alpha_{\rm Fe}$ for
Ba$_{0.6}$K$_{0.4}$Fe$_2$As$_2$, SmFeAsO$_{0.85}$F$_{0.15}$, and FeSe$_{1-x}$
with the corresponding values $\alpha_{\rm Fe}=0.34(3)$, 0.37(3), and 0.81(15),
respectively. Note that $\alpha_{\rm Fe}=0.81(15)$ for FeSe$_{1-x}$ exceeds
grossly the BCS value. In the other two studies Shirage {\it et al.} have
reported a {\it negative} $\alpha_{\rm Fe}=-0.18(3)$ and $-0.024(15)$ for
Ba$_{0.6}$K$_{0.4}$Fe$_2$As$_2$ \cite{Shirage09} and
SmFeAsO$_{1-y}$,\cite{Shirage10} respectively.
These controversial results are unlikely to stem from different pairing
mechanisms to be realized in different Fe-based superconductors. Especially, in
the case of Ba$_{0.6}$K$_{0.4}$Fe$_2$As$_2$, nominally identical samples were
isotope replaced with one exhibiting a positive\cite{Liu09} and the other a
negative isotope exponent.\cite{Shirage09} Note, that the sign reversed isotope
exponent seen by Shirage {\it et al.}\cite{Shirage09,Shirage10} was attributed
to multi-band superconductivity with different pairing channels, namely a
phononic one and an antiferromagnetic (AF) fluctuation dominated
one.\cite{Yanagisawa09} On the other hand a multi-band model cannot exhibit
{\it any} sign reversed isotope exponent even if solely AF fluctuations were
the pairing glue.\cite{Bussmann-Holder10}

In the present study we demonstrate that the very controversial results for
$\alpha_{\rm Fe}$ are caused by small structural changes occurring
simultaneously with the Fe isotope exchange. As such, we decompose the Fe-IE
exponent into one related to the structural changes $\alpha_{\rm Fe}^{\rm str}$
and the genuine (intrinsic)  one $\alpha_{\rm Fe}^{\rm int}$ to arrive at:
\begin{equation}
\alpha_{\rm Fe}=\alpha_{\rm Fe}^{\rm int}+\alpha_{\rm Fe}^{\rm str}.
 \label{eq:alpha-tot}
\end{equation}
By comparing the c-axis lattice constants for the pairs of isotopically
substituted samples we observe that $\alpha_{\rm Fe}^{\rm str}$ is negative for
Ba$_{0.6}$K$_{0.4}$Fe$_2$As$_2$ and SmFeAsO$_{1-y}$ studied by Shirage {\it et
al.} in Refs.~\onlinecite{Shirage09} and \onlinecite{Shirage10}, positive for
FeSe$_{1-x}$ from Ref.~\onlinecite{Khasanov10_FeSe-isotope} and close to 0 for
Ba$_{0.6}$K$_{0.4}$Fe$_2$As$_2$ and SmFeAsO$_{0.85}$F$_{0.15}$ measured in
Ref.~\onlinecite{Liu09}.
By taking into account the sign of $\alpha_{\rm Fe}^{\rm str}$ we arrive at the
conclusion that $\alpha_{\rm Fe}^{\rm int}$ is positive for all so far studied
Fe-based HTS.

\begin{figure}[htb]
%\centering
\includegraphics[width=1.0\linewidth]{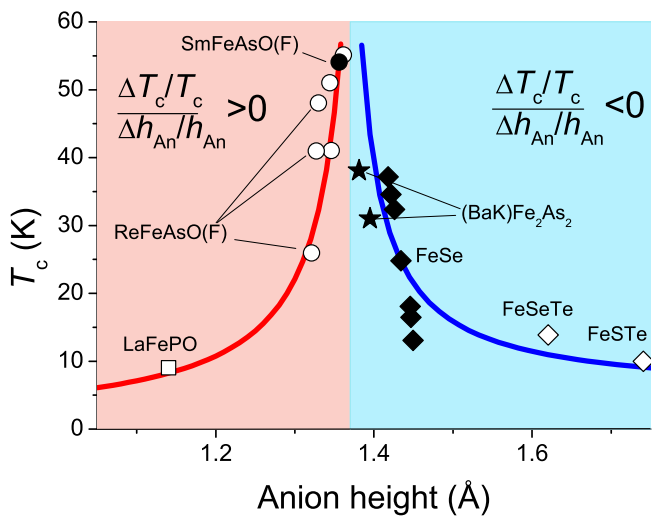}
% \vspace{-1.0cm}
%
\caption{(color online) Dependence of the superconducting transition
temperature $T_c$ on the height of the anion atom ($h_{\rm An}$, An=As, Se, P)
for varios families of Fe-based HTS, after Mizuguchi {\it et
al.}\cite{Mizuguchi10}). The closed symbols represent the samples which are
relevant for the present study. The lines are guided for the eye. The red(blue)
area represents part of $T_c$ vs. $h_{\rm An}$ diagram where $T_c$
increases(decreases) with increasing $h_{\rm An}$. }
 \label{fig:Pnictogen-height}
\end{figure}

\begin{table*}[htb]
\caption[~]{\label{Table:IE-results}  Summary of Fe isotope effect studies for
FeSe$_{1-x}$ Ref.~\onlinecite{Khasanov10_FeSe-isotope},
SmFeAsO$_{0.85}$F$_{0.15}$ and Ba$_{0.6}$K$_{0.4}$Fe$_2$As$_2$
Ref.~\onlinecite{Liu09}, Ba$_{0.6}$K$_{0.4}$Fe$_2$As$_2$
Ref.~\onlinecite{Shirage09} and SmFeAsO$_{1-x}$ Ref.~\onlinecite{Shirage10}.
The parameters are: $T_c$ -- superconducting transition tempreature for the
sample with the natural Fe isotope ($^{\rm nat}$Fe); $\alpha_{\rm Fe}$ -- Fe
isotope effect exponent; c -- the c-axis lattice constant for the sample with
the lighter ($^{\rm light}$Fe) and the heavier ($^{\rm heavy}$Fe) Fe isotope;
$\Delta c/c$ -- the relative shift of the c-axis constant caused by the Fe
isotope substitution; $\alpha_{\rm Fe}^{\rm str}$ and $\alpha_{\rm Fe}^{\rm
int}$ -- the structural and the intrinsic contributions to $\alpha_{\rm Fe}$.
See text for details.}
\begin{center}
% \vspace{-0.5cm}
\begin{tabular}{cccccccccccc}\\
 \hline
 \hline
%
%
%&& \multicolumn{3}{c}{Light Fe isotope}&&& \multicolumn{3}{c}{Heavy Fe isotope}&&\\
%
%
Sample&Reference&$T_c(^{\rm nat}$Fe)&$\alpha_{\rm Fe}$&c-axis($^{\rm
light}$Fe)&c-axis($^{\rm heavy}$Fe)&$\Delta {\rm c}/{\rm c}$
&$\alpha_{\rm Fe}^{\rm str}$&$\alpha_{\rm Fe}^{\rm int}$\\
&&(K)&&(\AA)&(\AA)&&&\\
\hline
FeSe$_{1-x}$ & Ref.~\onlinecite{Khasanov10_FeSe-isotope}&8.21(4) &0.81(15) &5.48683(9)&5.48787(9)&$>0$&$\simeq0.4$&$\simeq0.4$ \\
Ba$_{0.6}$K$_{0.4}$Fe$_2$As$_2$&Ref.~\onlinecite{Liu09}&37.30(2) &0.37(3) &13.289(7)&13.288(7)&$\sim0$&$\sim0$&$\sim0.35$ \\
Ba$_{0.6}$K$_{0.4}$Fe$_2$As$_2$&Ref.~\onlinecite{Shirage09}&37.78(2) &$-0.18(3)$ &13.313(1)&13.310(1)&$<0$&$\sim-0.5$&--\\
SmFeAsO$_{0.85}$F$_{0.15}$&Ref.~\onlinecite{Liu09}&41.40(2) &0.34(3) &8.490(2)&8.491(2)&$\sim0$&$\sim0$&$\sim0.35$ \\
SmFeAsO$_{1-y}$&Ref.~\onlinecite{Shirage10}&54.02(13) &$-0.024(15)$ &8.4428(8)&8.4440(8)&$\gtrsim0$&$<0$& --\\
 \hline \hline \\

\end{tabular}
   \end{center}
\end{table*}

Our motivation to separate the isotope coefficient into the above mentioned two
components, see Eq.~(\ref{eq:alpha-tot}),  stems from the fact that
superconductivity in these compounds is intimately related to small structural
changes as reported in various works. As an example, we mention the strong
nonlinear dependence of the superconducting transition temperature on the anion
atom height ($h_{\rm An}$, An=As, P, or Se) with a sharp maximum of $T_c$ at
$h_{\rm An}\simeq1.38$~\AA, see Mizuguchi {\it et al.}\cite{Mizuguchi10} and
Fig.~\ref{fig:Pnictogen-height}. The influence of the Fe isotope substitution
on the crystal structure, on the other hand, was first considered by Granath
{\it et al.}\cite{Granath09} based on the results of Raman studies of
CaFeAsO$_{1-x}$ and NdFeAsO$_{1-x}$ and further confirmed by Khasanov {\it et
al.}\cite{Khasanov10_FeSe-isotope} in neutron powder diffraction experiments on
$^{54}$Fe to $^{56}$Fe substituted FeSe$_{1-x}$.

The c-axis lattice constants for the pairs of Fe isotope substituted samples
Ba$_{0.6}$K$_{0.4}$Fe$_2$As$_2$, SmFeAsO$_{0.85}$F$_{0.15}$, SmFeAsO$_{1-y}$,
and FeSe$_{1-x}$  studied in
Refs.~\onlinecite{Liu09,Shirage09,Khasanov10_FeSe-isotope,Shirage10} are
summarized in Table~\ref{Table:IE-results}. The choice of the c-axis lattice
constant as the relevant quantity in deriving the structural isotope effect
might appear to be rather arbitrary since $T_c$ is influenced by all structural
details, namely tetrahedral angle, a-axis lattice constant, internal bond
lengths, {\it etc}. However, the c-axis lattice constant provides a very
sensitive probe since its compression(expansion) is directly accompanied by the
corresponding variation of the distance from the Fe-planes to the above(below)
lying anions\cite{Margadonna09,Rotter08,McQueen09} which, in turn, is a well
characterized property for many Fe-based compounds.\cite{Mizuguchi10}
From Table ~\ref{Table:IE-results} it is obvious that in
Ba$_{0.6}$K$_{0.4}$Fe$_2$As$_2$ and SmFeAsO$_{0.85}$F$_{0.15}$\cite{Liu09} the
c-axis constants are the same within the experimental error for both
isotopically substituted sets of the samples.\cite{comment} In FeSe$_{1-x}$
\cite{Khasanov10_FeSe-isotope} the c-axis constant is larger, while in
Ba$_{0.6}$K$_{0.4}$Fe$_2$As$_2$ \cite{Shirage09} it is smaller for the sample
with the heavier Fe isotope. In SmFeAsO$_{1-y}$, studied by Shirage {\it et
al.},\cite{Shirage10} both c-axis lattice constants seem to coincide within the
experimental resolution. However, since the difference between them is 1.5
times larger than one standard deviation, it is conceivable to attribute an
increase in the c-axis lattice constant in SmFeAsO$_{1-y}$ with the heavier Fe
isotope.

\begin{figure}[htb]
%\centering
\includegraphics[width=1.0\linewidth]{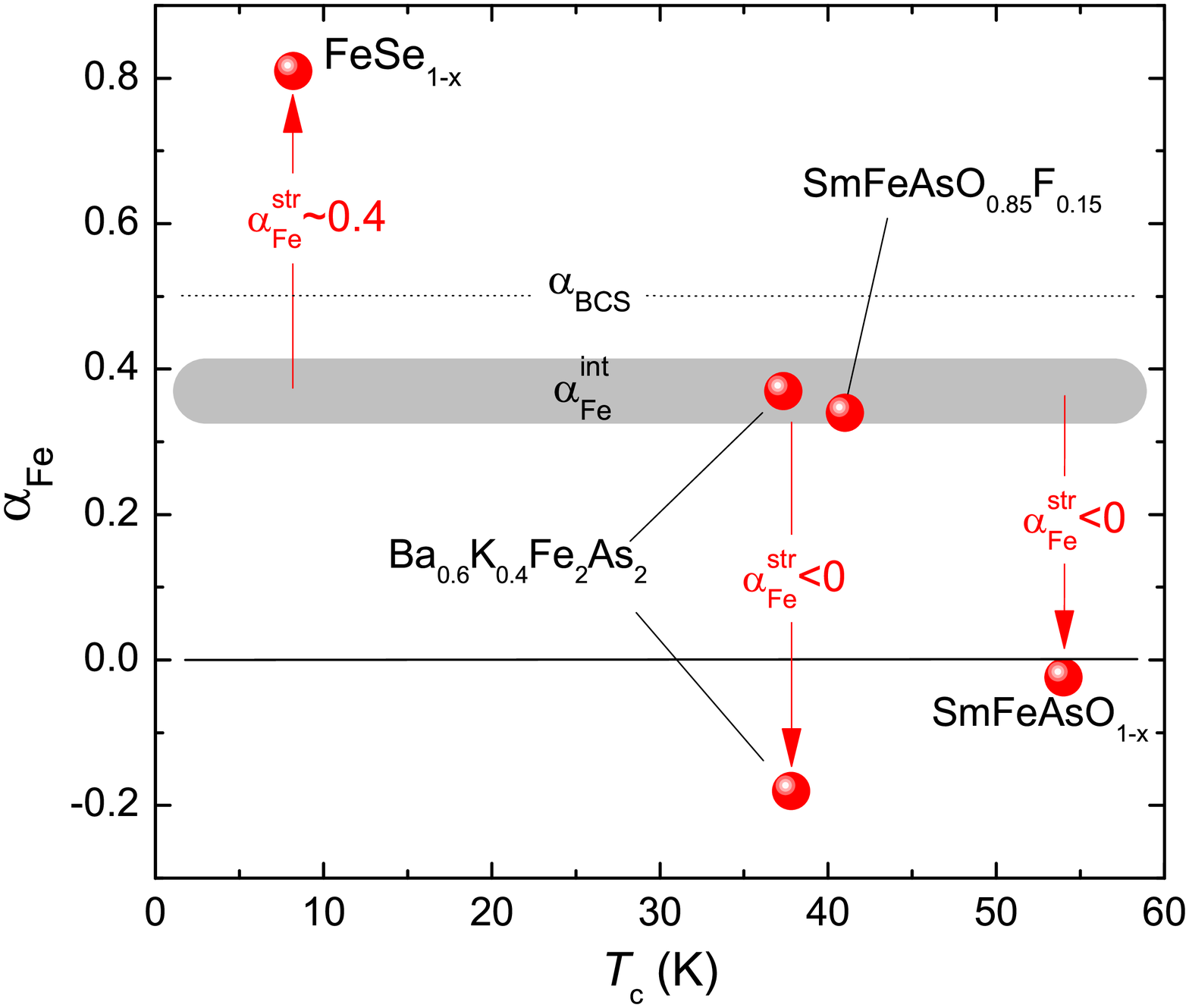}
% \vspace{-1.0cm}
%
\caption{(color online) Fe isotope effect exponent $\alpha_{\rm Fe}$ as a
function of the superconducting transition temperature $T_c$ for the samples
considered in the present study: FeSe$_{1-x}$
Ref.~\onlinecite{Khasanov10_FeSe-isotope}, Ba$_{0.6}$K$_{0.4}$Fe$_2$As$_2$ and
SmFeAsO$_{0.85}$F$_{0.15}$ Ref.~\onlinecite{Liu09},
Ba$_{0.6}$K$_{0.4}$Fe$_2$As$_2$ Ref.~\onlinecite{Shirage09}, and
SmFeAsO$_{1-x}$ Ref.~\onlinecite{Shirage10}. Arrows indicate the direction of
the shift from the ``intrinsic'' Fe-isotope effect exponent $\alpha_{\rm
Fe}^{\rm int}\sim0.35-0.4$ caused by the structural effects. $\alpha_{\rm
BCS}\equiv0.5$ is the BCS value for electron-phonon mediated superconductivity.
See text for details }
 \label{fig:Isotope-exponent}
\end{figure}

The use of the empirical $T_c$ vs. $h_{\rm An}$ relation from
Ref.~\onlinecite{Mizuguchi10} combined with the intrinsic relation of the
proportionality between the c-axis constant and the anion atom height
($\Delta{\rm c}\propto\Delta h_{\rm An}$, see
Refs.~\onlinecite{Margadonna09,Rotter08,McQueen09}) enables us to determine the
sign of the structurally related shift of $T_c$  induced by isotopic exchange.
By defining the shift of a given quantity $X$ as $\Delta X/X=(\;^{^{\rm
light}{\rm Fe}}X-\;^{^{\rm heavy}{\rm Fe}}X)/\;^{^{\rm heavy}{\rm Fe}}X$ and
following Mizuguchi {\it et al.}\cite{Mizuguchi10},  see also
Fig.~\ref{fig:Pnictogen-height}, the sign of $(\Delta T_c/T_c)/(\Delta h_{\rm
An}/h_{\rm An})$ is positive for SmFeAsO(F) as well as for various Fe-based HTS
belonging to ReFeAsO(F) family (Re=Nd, Ce, La) and negative for
(BaK)Fe$_2$As$_2$ and FeSe$_{1-x}$. Consequently the change of the c-axis
constant caused by Fe isotope substitution as presented in
Table~\ref{Table:IE-results} results in an additional structurally related
shift of $T_c$ being positive for FeSe$_{1-x}$,\cite{Khasanov10_FeSe-isotope}
negative for Ba$_{0.6}$K$_{0.4}$Fe$_2$As$_2$ and
SmFeAsO$_{1-y}$,\cite{Shirage09,Shirage10} and close to 0 for
Ba$_{0.6}$K$_{0.4}$Fe$_2$As$_2$ and SmFeAsO$_{0.85}$F$_{0.15}$.\cite{Liu09} It
is rather remarkable that the corresponding ``structural'' Fe-IE exponents
$\alpha_{\rm Fe}^{\rm str}$ would lead to the shift of genuine (intrinsic)
$\alpha_{\rm Fe}^{\rm int}$ in the direction of 0.35--0.4, see
Fig.~\ref{fig:Isotope-exponent}.

Note that the above mentioned discussion allows only to determine the sign of
the structurally related isotope effect but not the absolute value. The reasons
are the following. First, the relative change of the c-axis constant is
proportional, but not identical to the one of $h_{\rm An}$. As an example,
$^{56}$Fe to $^{54}$Fe isotope substitution in FeSe$_{1-x}$ leads to an
increase of the c-axis constant by approximately 0.02\%, while the change of
the Se height amounts to $\simeq0.22$\%, see
Ref.~\onlinecite{Khasanov10_FeSe-isotope}. Second, the height of the anion atom
is clearly not the only parameter which is crucial for $T_c$ of Fe-based HTS as
already mentioned above. However, the lack of a consistent structural
characterization limits this study to a single parameter which was emphasized
to be of uppermost relevance to $T_c$.

The analysis of the structural data together with the dependence of $T_c$ on Se
height in FeSe$_{1-x}$, as well as on Se(Te)--Fe--Se(Te) angle and the a-axis
constant in FeSe$_{1-y}$Te$_y$ for $y\leq0.5$ admits to extract the
``structural'' Fe isotope effect exponent $\alpha_{\rm Fe}^{\rm str}\simeq 0.4$
for $^{56}$Fe to $^{54}$Fe substituted FeSe$_{1-x}$
samples.\cite{Khasanov10_FeSe-isotope} The absence of precise structural data
complicates the precise analysis as outlined in
Refs.~\onlinecite{Liu09,Shirage09,Shirage10}.
However, a zero, within the experimental accuracy, Fe isotope shift of the
c-axis lattice constant for Ba$_{0.6}$K$_{0.4}$Fe$_2$As$_2$ as reported by Liu
{\it et al.}\cite{Liu09} is a clear indication that no structural effect is
present for this particular set of the samples. Consequently, the negative
isotope effect exponent $\alpha_{\rm Fe}\simeq-0.18$ obtained for nominally
identically doped Ba$_{0.6}$K$_{0.4}$Fe$_2$As$_2$ by Shirage {\it et
al.}\cite{Shirage09} stems from summing both effects, {\it i.e.},
$-0.18(\alpha_{\rm Fe})=0.35(\alpha_{\rm Fe}^{\rm int} )-0.52(\alpha_{\rm
Fe}^{\rm str} )$, see Eq.~(\ref{eq:alpha-tot}).
It is important to recognize that for SmFeAsO(F) a similar analysis is
impossible since samples with different doping levels (different $T_c$'s, see
Table~\ref{Table:IE-results} and Fig.~\ref{fig:Isotope-exponent}) were studied
in Refs.~\onlinecite{Liu09} and \onlinecite{Shirage10}.

Recently Bussmann-Holder {\it et al.}\cite{Bussmann-Holder09} investigated a
multiple gap scenario of superconductivity in Fe-based HTS with the aim to
search for possible sources of the isotope effect on $T_c$. Typical phonon
mediated scenarios were contrasted to polaronic effects and found to have very
different impacts on the isotope effect. While phonon mediated
superconductivity slightly suppresses the isotope effect as compared to the BCS
value $\alpha_{\rm BCS}\equiv0.5$, polaronic effects can largely enhance it.
The scenario of electron-phonon mediated superconductivity within the dominant
gap channel predicts a $T_c$ independent isotope effect with the $\alpha$ value
being slightly smaller than 0.5 thus  agreeing rather well with that observed
for FeSe$_{1-x}$,\cite{Khasanov10_FeSe-isotope}
Ba$_{0.6}$K$_{0.4}$Fe$_2$As$_2$,\cite{Liu09,Shirage09} and
SmFeAsO$_{0.85}$F$_{0.15}$.\cite{Liu09} Indeed, for these particular samples,
which belong to 3 different families of Fe-based HTS and have $T_c$'s between 8
and 44~K, the ``intrinsic'' Fe isotope exponent is almost constant with
$\alpha_{\rm Fe}^{\rm int}\sim0.35-0.4$, see Table~\ref{Table:IE-results} and
Fig.~\ref{fig:Isotope-exponent}.
As such, the independent on $T_c$ value of $\alpha_{\rm Fe}^{\rm int}$ would
suggest $\alpha_{\rm Fe}^{\rm str}\sim-0.4$ for SmFeAsO$_{1-x}$ studied by
Shirage {\it et al.}\cite{Shirage10}

To conclude, the currently available Fe isotope effect data on the
superconducting transition temperature $T_c$ for various Fe-based HTS were
reanalyzed by separating the measured Fe-IE exponent $\alpha_{\rm Fe}$ into a
structural and an intrinsic (unrelated to the structural changes) component.
Accounting for the empirical relation between $T_c$ and the anion atom height
$h_{\rm An}$\cite{Mizuguchi10} we have demonstrated that the structural
contribution to the Fe-IE exponent is negative for
Ba$_{0.6}$K$_{0.4}$Fe$_2$As$_2$ and SmFeAsO$_{1-x}$ studied by Shirage {\it et
al},\cite{Shirage09,Shirage10} positive for
FeSe$_{1-x}$,\cite{Khasanov10_FeSe-isotope} and close to 0 for
SmFeAsO$_{0.85}$F$_{0.15}$ and Ba$_{0.6}$K$_{0.4}$Fe$_2$As$_2$ measured by Liu
{\it et al.}\cite{Liu09} By taking such corrections into account we infer that
the value of the genuine  Fe-IE exponent is close to $\alpha_{\rm Fe}^{\rm
int}\sim 0.35-0.4$ for compounds belonging to at least three different families
of Fe-based HTS. We are convinced that the analysis presented in our paper
helps in clarifying the existing controversy on the isotope effect in Fe-based
superconductors.


\begin{thebibliography}{99}
%
\bibitem{Poole00} C.P.~Poole, {\it Hnadbook of Superconductitivy}, Academic
press, 24-28 Oval Road, London, (2000).
%
\bibitem{IE-PdH} W.~Buckel and B.~Strizker, Phys.~Letters {\bf 43A}, 403
(1973);  J.E.~Schriber and C.J.M.~Northrup, Jr., Phys.~Rev.~B {\bf 10}, 3818
(1974).
%
\bibitem{Yussouff95} M.~Yussouff, B.K.~Rao, and P.~Jena Sol.~State.~Commun. {\bf 94}, 549
(1995).
%
\bibitem{Rahman76} A.~Rahman, K.~Sk\"{o}ld, C.~Pelizzari, S. K. Sinha, and H.~Flotow,
Phys.~Rev.~B {\bf 14}, 3630 (1976).
%
\bibitem{Schlueter01} J.A.~Schlueter, A.M.~Kini, B.H.~Ward, U.~Geiser, H.H.~Wang, J.~Mohtasham, R.W.~Winter and G.L.~Gard,
Physica~C {\bf 351}, 261 (2001).
%
\bibitem{Whangbo97} M.H.~Whangbo, J.M.~Willimas, A.J.~Schultz, T.J.~Emge, and
M.A.~Beno, J.~Am.~Chem.Soc., 109, 90 (1997).
%
\bibitem{Batlogg87} B.~Batlogg, G.~Kourouklis, W.~Weber, R.J.~Cava,
A.~Jayaraman, A.E.~White, K.T.~Short, L.W.~Rupp, and E.A.~Rietman,
Phys.~Rev.~Lett. {\bf 59}, 912 (1987).
%
\bibitem{Franck91Franck94}J.P.~Franck, J.~Jung,  M.A-K.~Mohamed, S.~Gygax, and G.I.~Sproule,
Phys.~Rev.~B {\bf 44}, 5318 (1991); J.P.~Franck,  in {\it Physical Properties
of High Temperature Superconductors IV}, edited by D.~M.~Ginsberg (World
Scientific, Singapore, 1994), pp.~189--293.
%
\bibitem{Zech94} D.~Zech, H.~Keller, K.~Conder, E.~Kaldis, E.~Liarokapis,
N.~Poulakis, and K.A.~M\"uller, Nature~(London) {\bf 371}, 681 (1994).
%
\bibitem{Zhao01} G.-M.~Zhao, H.~Keller, and K.~Conder, J.~Phys.:~Condens.~Matter {\bf 13}, R569
(2001).
%
\bibitem{Khasanov04Khasanov04aKhasanov06Khasanov07Khasanov08} R.~Khasanov, A.~Shengelaya,
E.~Morenzoni, K.~Conder, I.M.~Savi\'c, and H.~Keller,
J.~Phys.:~Condens.~Matter {\bf 16}, S4439 (2004);
%
R.~Khasanov, S.~Str\"assle, K.~Conder, E.~Pomjakushina, A.~Bussmann-Holder, and
H.~Keller, Phys.~Rev.~B {\bf 77}, 104530 (2008).
%
\bibitem{Tallon05} J.L.~Tallon, R.S.~Islam, J.~Storey, G.V.M.~Williams, and J.R.~Cooper,
Phys.~Rev.~Lett. {\bf 94}, 237002 (2005).
%
\bibitem{IE_Bi2201-2212-2223} X.-J.~Chen, B.~Liang, C.~Ulrich, C.-T.~Lin, V.V.~Struzhkin,
Z.~Wu, R.J.~Hemley, H.-k.~Mao, and H.-Q.~Lin, Phys.~Rev.~B {\bf 76}, 140502
(2007).
%
\bibitem{Khasanov08_IE-phase-diagram} R.~Khasanov, A.~Shengelaya, D.~Di~Castro, E.~Morenzoni,
A.~Maisuradze, I.M.~Savic, K.~Conder, E.~Pomjakushina, A.~Bussmann-Holder, and
H.~Keller, Phys.~Rev.~Lett. {\bf 101}, 077001 (2008).
%
%
\bibitem{Liu09} R.H.~Liu, T.~Wu, G.~Wu, H.~Chen, X.F.~Wang, Y.L.~Xie,
J.J.~Yin, Y.J.~Yan, Q.J.~Li, B.C.~Shi, W.S.~Chu, Z.Y.~Wu, and X.H.~Chen, Nature
{\bf 459}, 64 (2009).
%
\bibitem{Shirage09} P.M.~Shirage, K.~Kihou, K.~Miyazawa, C.-H.~Lee, H.~Kito, H.~Eisaki,
T.~Yanagisawa, Y.~Tanaka, and A.~Iyo, Phys.~Rev.~Lett {\bf 103}, 257003 (2009).
%
\bibitem{Shirage10} P.M.~Shirage, K.~Miyazawa, K.~Kihou, H.~Kito,
Y.~Yoshida, Y.~Tanaka, H.~Eisaki, and A.~Iyo, Phys.~Rev.~Lett.
{\bf 105}, 037004 (2010).
%
\bibitem{Khasanov10_FeSe-isotope} R.~Khasanov, M.~Bendele, K.~Conder, H.~Keller, E.~Pomjakushina, and
V.~Pomjakushin, New~J.~Phys. {\bf 12}, 073024 (2010).
%
\bibitem{Yanagisawa09} T.~Yanagisawa, K.~Odagiri, I.~Hase, K.~Yamaji, P.M.~Shirage, Y.~Tanaka,
A.~Iyo, and H.~Eisaki, J.~Phys.~Soc.~Jpn. {\bf 78}, 094718 (2009).
%
\bibitem{Bussmann-Holder10} A.~Bussmann-Holder and H.~Keller, unpublished.
%
\bibitem{Mizuguchi10} Y.~Mizuguchi, Y.~Hara, K.~Deguchi, S.~Tsuda, T.~Yamaguchi,
K.~Takeda, H.~Kotegawa, H.~Tou, and Y.~Takano, Supercond.~Sci.~Technol. {\bf
23} 054013 (2010).
%
\bibitem{Granath09} M.~Granath, J.~Bielecki, J.~Holmlund, and L.~B\"{o}rjesson, Phys.~Rev.~B {\bf 79},
235103 (2009).
%
\bibitem{Margadonna09} S.~Margadonna, Y.~Takabayashi, M.T.~McDonald, M.~Brunelli, G.~Wu, R.H.~Liu, X.H.~Chen, and
K.~Prassides, Phys.~Rev.~B {\bf 79}, 014503 (2009).
%
\bibitem{Rotter08} M.~Rotter, M.~Pangerl, M.~Tegel, and D.~Johrendt, Angew.~Chem.~Int.~Ed. {\bf
47}, 7949 (2008).
%
\bibitem{McQueen09} T.M.~McQueen, Q.~Huang, V.~Ksenofontov, C.~Felser, Q.~Xu,
H.W.~Zandbergen, Y.S.~Hor, J.~Allred, A.J.~Williams, D.~Qu, J.~Checkelsky,
N.P.~Ong, and R.J.~Cava, Phys.~Rev.~B {\bf 79}, 014522 (2009).
%
\bibitem{comment} The c-axis lattice constants in isotopically substituted samples are the same
within $\sim1\times10^{-3}$~\AA, while the absolute errors are  larger by the
factor of 7 for SmFeAsO$_{85}$F$_{0.15}$ and by the factor of 2 for
Ba$_{0.6}$K$_{0.4}$Fe$_2$As$_2$.
%
%\bibitem{Zhao08} J.~Zhao, Q.~Huang, C.~de~la~Cruz, S.~Li, J.W.~Lynn, Y.~Chen, M.A.~Green, G.F.~Chen, G.~Li,
%Z.~Li, J.L.~Luo, N.L.~Wang, and P.~Dai, Nature Materials {\bf 7}, 953 (2008).
%
\bibitem{Bussmann-Holder09} A.~Bussmann-Holder, A.~Simon, H.~Keller, and
A.R.~Bishop, arXiv:0906.2283.
%
\end{thebibliography}
\end{document}